\documentclass[prl,twocolumn,showpacs,preprintnumbers,amsmath,amssymb]{revtex4}

\usepackage{graphicx}
\usepackage{dcolumn}
\usepackage{bm}
\begin{document}


\title{Physics, Stability and Dynamics of Supply Networks}
\author{Dirk Helbing, Stefan L\"ammer, and Thomas Seidel}
  \email{helbing@trafficforum.org}
  \homepage{http://www.helbing.org}
\affiliation{Dresden University of Technology, Andreas-Schubert-Str. 23, 
01069 Dresden, Germany}

\author{P\'{e}tr \v{S}eba}
\affiliation{Institute of Physics, Czech Academy of Science, Cukrovarnick\'a 10, 
162 53 Prague, Czech Republic}

\author{Tadeusz P\l{}atkowski}
\affiliation{Department of Mathematics, Informatics and Mechanics, 
University of Warsaw, Banacha 2, 
02-097 Warsaw, Poland}

\date{\today}


\begin{abstract}
We show how to treat supply networks as physical transport problems governed by balance
equations and equations for the adaptation of production speeds. Although the non-linear behaviour
is different, the linearized set of coupled differential equations is formally
related to those of mechanical or electrical oscillator networks. Supply networks possess interesting new
features due to their complex topology and directed links. We derive analytical conditions for absolute
and convective instabilities. The empirically observed ``bull-whip effect'' in supply chains is
explained as a form of convective instability based on resonance effects. Moreover, it is generalized
to arbitrary supply networks. Their related eigenvalues are usually complex, depending on the
network structure (even without loops). Therefore, their generic behavior is characterized by oscillations.
We also show that regular distribution networks possess two negative
eigenvalues only, but perturbations generate a spectrum of complex eigenvalues.
\end{abstract}
\pacs{89.65.Gh,
89.75.Hc,
47.70.-n,
84.30.Bv}
\maketitle

Econophysics has stimulated a lot of interesting research on problems in finance
and economic systems \cite{Mant1St99}, applying methods from statistical physics and the theory of
complex systems. Recently, the dynamics of supply networks \cite{Radons,NJP,stabilization}, i.e.
the flow of materials through networks, has been identified as an interesting
physical transport problem \cite{NJP,stabilization,Dag},  which is also reflected by the
notion of ``factory physics'' \cite{FactPhys}. 
Potential applications reach from production networks to business cycles,
from metabolic networks to food webs, up to logistic problems in disaster management.
Empirical and theoretical studies 
have shown that the flow of goods between different producers or suppliers can be described
analogously to driven many-particle flows between sources and sinks (depots), where the particles represent 
materials, goods, or other resources. In contrast to the stationary behavior assumed by most production engineers, 
this flow may display a complex dynamics in time, including oscillatory patterns and chaos \cite{Chaos,beer1}. 
A particular focus has been on the empirically observed and well-known 
{\em ``bull-whip effect''} \cite{Dag,beer1,Fluid,bull,NJP}, which describes the amplification
of the oscillation amplitudes of delivery rates in supply chains compared to the variations in
the consumption rate of goods. 
\par
A promising approach to the non-linear interactions and dynamics
of supply chains is based on ``fluid-dynamic models'' \cite{Fluid,Dag}, which are related 
to macroscopic traffic models \cite{NJP,stabilization,Dag}. In contrast to classical approaches like queueing theory and
event-driven simulations, they are better suited for an on-line control under dynamically changing conditions. 
These fluid-dynamic models have recently been generalized to cope with discrete 
spaces (production steps) \cite{NJP,Dag}. However, the impact of the topology of supply networks, 
connecting the subject to the statistical physics of networks \cite{netz} has not yet been thoroughly 
investigated. Its relevance for the stability and dynamics of supply networks will, therefore, 
be the main focus of this paper. Here, we will specifically concentrate on analytical results for the dynamic
behavior in the linear regime around the usually assumed stationary state, while non-linear effects are investigated
elsewhere \cite{NJP}. 
The corresponding equations can be mapped onto the ones of particular mechanical or electrical networks,
but supply networks are generally more complex and their non-linear dynamics is different. 
In particular, supply networks are directed and possess other characteristic topologies \cite{NATU}. Hence, 
our analytical investigation yields interesting new results. 
Apart from a generalization of the bull-whip effect from
linear (sequential) supply chains to networks, they allow us to explain various surprising features 
of numerical simulations of supply networks: (i) The bull-whip effect can occur even though the eigenvalues 
of linear supply chains are negative. (ii) Supply {\em networks} tend to show oscillations, even without 
loops in the material flows. (iii) Regular distribution networks are characterized by two negative
eigenvalues, but random perturbations in the network structure cause a spectrum of oscillating modes.
\par
Our simplified model of supply networks consists of $u$ suppliers $i$ delivering products to
other suppliers $j$ or consumers. The suppliers $j$ (e.g. production units of a plant or company)
will be assumed to deliver products, resources or materials
$k$ at a certain rate $d_{kj}X_j(t)$. Their demand of goods of kind $i$ 
from other suppliers $j$ is given by $c_{ij}X_j(t)$, where $0 \le c_{ij} \le 1$. 
The proportionality $c_{ij}X_j(t)$ to the production rate $X_j(t)$ of products $j$ 
reflects that a quantity $c_{ij}$ of product $i$ is needed to
produce one unit of product $j$. The coefficients $c_{ij}$ define an input matrix $\mathbf{C} = (c_{ij})$
and $d_{ij}$ an output matrix $\mathbf{D}$. The interactions 
among the different suppliers $i$ and $j$ are represented by the supply matrix
$\mathbf{S}=\mathbf{D}-\mathbf{C}$. In addition, we need to consider the
inventories, stock levels, or availability $N_i(t)$ of products, materials, or resources $i$. 
\par
Our model for the dynamics of supply networks consists of two sets of equations, one for
the change of inventories $N_i(t)$ with time $t$ and another one for the adaptation of
the delivery or production rates $X_j(t)$ of the suppliers or production units 
$j$. The inventories $N_i(t)$ of goods of kind $i$
are described by material balance equations for the flows of goods, which could be viewed as a
discontinuous version of the continuity equation:
\begin{equation}
 \frac{dN_i}{dt} = \overbrace{\underbrace{\sum_{k=1}^u d_{ik} X_k(t)}_{\rm supply}}^{\rm inflow}  
- \underbrace{\bigg[ \overbrace{\sum_{j=1}^u c_{ij} X_j(t)}^{\rm re-entrant} 
+ \overbrace{Y_i(t)}^{\rm outflow} \bigg]}_{\rm demand} 
\label{conserv}
\end{equation}
with $0 \le c_{ij}, d_{ij} \le 1$  ($1 \le i, j \le u$) and the ``normalization conditions''
$d_{i0} = 1 - \sum_{k=1}^u d_{ik}  \ge 0$,  $c_{i,u+1} = 1 - \sum_{j=1}^u c_{ij} \ge 0$. 
Frequently, one assumes $d_{ij} = \delta_{ij}$ with $\delta_{ij} = 1$ for $i=j$ and $\delta_{ij} = 0$ otherwise \cite{NJP}.
This corresponds to situations in which production units or suppliers are defined such that they deliver
one kind of good only.  The quantity
\begin{equation}
 Y_i(t) = \underbrace{c_{i,u+1}X_{u+1}(t)}_{\rm consumption\ and\ losses} \quad
 - \underbrace{d_{i0} X_0(t)}_{\rm inflow\ of\ resources}  
\end{equation}
comprises the consumption rate of goods $i$, losses, and waste (the ``export'' of material),
minus the inflows into the considered system (the ``imports''). 
\par
Changes in the demand $Y_i(t)$ sooner or later require an adaptation of the production rates $X_j(t)$.
This adaptation takes time (to determine the change in demand, to finish the planning process
and administrative steps, and to adapt orders and production capacities). We will reflect this by
an adaptation time $T$ and assume that the change in the production rate is proportional to
the deviation of the actual production rate $X_j$ from the desired one $W_j(\{N_i\},\{dN_i/dt\})$,
which depends on the inventories $N_i$ and their changes $dN_i/dt$ in time:
$dX_j/dt = [ W_j(\{N_i\},\{dN_i/dt\} ) - X_j(t) ]/T$. Although a more general treatment is possible,
we will first focus on cases where the delivery rate $X_j$ of supplier $j$ is only 
adapted to the inventory $N_i$ of the dominant product $i$ given by $d_{ij} =\max_k d_{kj}$. Moreover, we will
enumerate suppliers $j$ according to their dominating products $i$, which implies $j=i$.
The resulting adaptation equation
\begin{equation}
 \frac{dX_i}{dt} = \frac{1}{T} \left[ W_i\left(N_i,dN_i/dt \right) - X_i(t) \right] 
\label{zwei}
\end{equation}
still implies an indirect dependence on the production and delivery rates $X_j$ of other suppliers 
$j\ne i$ via the dependence on $dN_i/dt$, see Eq.~(\ref{conserv}).
\par
For the desired production rate
$W_i(N_i,dN_i/dt)$, we will assume that 
(i) it is non-negative and decreasing with
increasing inventories, 
(ii) it is proportional to the steady-state inventory $\overline{N_i}$, 
(iii) it responds to the relative deviation $N_i /\overline{N_i}$ of 
the inventory $N_i$ from the steady-state inventory $\overline{N_i}$ and the
relative change $(dN_i/dt)/N_i(t)$ in time. This suggests the scaling relation
\begin{equation}
 W_i\left(N_i,\frac{dN_i}{dt}\right) 
\approx \nu \overline{N_i} P\left(\frac{N_i}{\overline{N_i}},\frac{1}{N_i}\,\frac{dN_i}{dt} \right) 
\label{scaling}
\end{equation}
with a scaling function $P$ which, for simplicity, is assumed to be identical for all sectors.
The stationary value $\overline{N_i}$ of the inventory $N_i$ can be easily determined by solving
the linear system of equations
\begin{equation}
\overline{Y}_i = \sum_{j=1}^u (d_{ij} - c_{ij}) \overline{X}_j  
= \nu P(1,0) \sum_{j=1}^u  (d_{ij} - c_{ij}) \overline{N}_j \, ,
\end{equation}
where $\overline{Y}_i$ denotes the time-average of the demand $Y_i(t)$, while
$\overline{X}_i$ and $\overline{N}_i$ denote the
stationary values corresponding to the case $Y_i = \overline{Y}_i$. By appropriate choice of $\nu$, 
we can set $P(1,0) = 1$.
\par 
Since we want to know whether the stationary state of the supply network is stable, as is often implicitly
assumed in queueing theory, we will focus on what happens in case of small deviations 
$x_i(\tau) = [X_i(\tau T) - \overline{X}_i]T$,  $n_i(\tau) = N_i(\tau T) - \overline{N}_i$, and
$y_i(\tau) = [Y_i(\tau T) - \overline{Y}_i]T$ from it. 
Here, we have introduced a scaling of the model variables to
dimensionless units. For example, $\tau = t/T$ is a dimensionless time, i.e. in the following
we will measure the time in units of $T$. Close to equilibrium, we can linearize
the model equations. In matrix notation they read
\begin{eqnarray}
 \frac{d}{d\tau} \left(
\begin{array}{c} \vec{n}(\tau) \\ \vec{x}(\tau) \end{array}\right)
&=& \mathbf{M} 
\left( \begin{array}{c} \vec{n}(\tau) \\ \vec{x}(\tau) \end{array}\right)
+ \left( \begin{array}{c} -\vec{y}(\tau)  \\ B\vec{y}(\tau) \end{array}\right) 
\label{matrix} \\
\mbox{with }
 \mathbf{M} &=& \left( \begin{array}{ccc}
\mathbf{0} & , & \mathbf{S} \\
-\mbox{$A$} \mathbf{E} & , & - \mathbf{E} - \mbox{$B$} \mathbf{S} 
\end{array}\right) \, .
\label{according}
\end{eqnarray}
Here, $A = -\nu T \, \partial P(1,0)/\partial z_1$ 
and $BT = - \nu T \, \partial P(1,0)/\partial z_2$ 
denote the negative values of the partial
derivatives of $T W_i(N_i,dN_i/dt)$ with respect to the first variable
$z_1=N_i/\overline{N_i}$ and the second one $z_2=(dN_i/dt)/N_i= (dN_i/d\tau)/(N_iT)$ 
in the equilibrium point $(N_i,dN_i/dt) = (\overline{N_i},0)$. 
The negative signs reflect that $A$ and $B$ are typically positive, as the production rate should decrease
when the inventory increases. 
\par
The above linear system of coupled differential equations is a quite general approach to the study of 
supply networks and material flows close to the stationary state. It describes the response of 
delivery rates to a variation $\vec{y}(t)$ in the consumption rate  and can be derived
from various non-linear supply chain models which differ in their degree of detailedness 
concerning the consideration of forecasts \cite{stabilization}, 
price dynamics \cite{NJP,NATU}, or lack of materials \cite{NJP}. Therefore, it is worth investigating 
the dependence on the dimensionless parameters $A$ and $B$, and on the structure of the supply
network characterized by $\mathbf{S}$. The system of $2u$ differential equations can also
be written as a system of $u$ second-order differential equations
\begin{equation}
 \frac{d^2\vec{x}}{d\tau^2} + [\mathbf{E} + \mbox{$B$}\mathbf{S}] \frac{d\vec{x}}{d\tau} 
+ \mbox{$A$}\mathbf{S} \vec{x}(\tau) = \vec{g}(t) \, , 
\label{dgl}
\end{equation}
where $\vec{g}(t) = A \vec{y}(\tau) + B \frac{d\vec{y}}{d\tau}$. Note that there exists a matrix
$\mathbf{T}$ which allows one to transform the matrix $\mathbf{E}- \mathbf{S}$
via $\mathbf{T}^{-1}(\mathbf{E} - \mathbf{S})\mathbf{T} = \mathbf{J}$ into either
a diagonal or a Jordan normal form $\mathbf{J}$. Defining $\vec{\mu}(\tau) = \mathbf{T}^{-1}\vec{x}(\tau)$
and $\vec{h}(\tau) = \mathbf{T}^{-1}\vec{g}(\tau)$, we obtain the coupled set of second-order
differential equations
\begin{equation}
 \frac{d^2\mu_i}{d\tau^2} + 2\gamma_i \frac{d\mu_i}{d\tau} + \omega_i{}^2 \mu_i 
=  b_i \left(B\mu_{i+1} + A \frac{d\mu_{i+1}}{d\tau} \right) + h_i  \, ,
\label{chain}
\end{equation}
where $\gamma_i = [1+B(1-J_{ii})]/2$, $\omega_i = [A(1-J_{ii})]^{1/2}$, and $b_i = J_{i,i+1}$.
This can be interpreted as a set of equations for linearly 
coupled damped oscillators with damping constants $\gamma_i$, 
eigenfrequencies $\omega_i$, and external forcing $h_i(\tau)$.
The other forcing terms on the right-hand side are due to interactions of suppliers. 
They appear only if $\mathbf{J}$ is not of diagonal, but of Jordan normal form. 
Because of $b_i = J_{i,i+1}$, Eqs. (\ref{chain}) can always be analytically
solved in a recursive way, starting with the highest index $i=u$.
Note that, in the case $\mathbf{D}=\mathbf{E}$ (i.e. $d_{ij}=\delta_{ij}$), $J_{ii}$ are the eigenvalues of
the input matrix $\mathbf{C}$ and $0 \le |J_{ii}| \le 1$. 
Equation~(\ref{chain}) has a special periodic solution of the form
$\mu_{i}(t) = \mu_{i}^0 \, \mbox{e}^{{\rm i}(\alpha t - \beta_{i})}$, 
$h_i(t) = h_i^0 \, \mbox{e}^{{\rm i}\alpha t}$, 
where ${\rm i} = \sqrt{-1}$ denotes the imaginary unit. Inserting this into (\ref{chain}) and dividing
by $\mbox{e}^{{\rm i}\alpha t}$ immediately gives
\begin{equation}
  (-\alpha^2 + 2{\rm i}\alpha \gamma_i + \omega_i{}^2) \mu_i^0 \mbox{e}^{-{\rm i}\beta_i}
 = b_i (A + {\rm i}\alpha B) \mu_{i+1}^0 \mbox{e}^{-{\rm i}\beta_{i+1}} + h_i^0 \, .
\end{equation}
With $\mbox{e}^{\pm{\rm i}\phi} = \cos (\phi) \pm {\rm i} \sin(\phi)$ this implies
\begin{equation}
 \mu_i^0 \mbox{e}^{-{\rm i}\beta_i} 
= \frac{\sqrt{\mbox{Re}^2+\mbox{Im}^2} \, \mbox{e}^{{\rm i}\rho_i}}
{\sqrt{(\omega_i{}^2 - \alpha^2)^2 + (2\alpha\gamma_i)^2} \, \mbox{e}^{{\rm i}\varphi_i}} \, ,
\end{equation}
where $\mbox{Re} = b_i\mu_{i+1}^0 [ A\cos(\beta_{i+1})+ \alpha B\sin(\beta_{i+1})] + h_i^0$ and
$\mbox{Im} = b_i\mu_{i+1}^0 [ \alpha B \cos(\beta_{i+1})- A \sin(\beta_{i+1})]$. 
\begin{equation}
 \mu_i^0 = \sqrt{ \frac{[A^2+(\alpha B)^2](b_i\mu_{i+1}^0)^2 + h_i^0 H_i + (h_i^0)^2 }  
{(\omega_i{}^2 - \alpha^2)^2 + (2\alpha\gamma_i)^2} } 
\label{help}
\end{equation}
with $H_i = 2 b_i \mu_{i+1}^0 [A \cos(\beta_{i+1}) + \alpha B \sin(\beta_{i+1})]$.
Finally, we have $\beta_i = \varphi_i - \rho_i$ with $\tan \varphi_i 
= 2\alpha \gamma_i/ (\omega_i{}^2 - \alpha^2)$ and
\begin{equation}
 \tan \rho_i = \frac{b_i\mu_{i+1}^0 [ \alpha B \cos(\beta_{i+1})- A \sin(\beta_{i+1})]}
{ b_i\mu_{i+1}^0 [A \cos(\beta_{i+1})+ \alpha B \sin(\beta_{i+1})] + h_i^0} \, .
\end{equation}
For $h_i^0 = 0$, we obtain $\tan (\varphi_i - \beta_i) = \tan(\delta - \beta_{i+1})$ with
$\tan \delta = \alpha B/A$, i.e. the phase shift between $i$ and $i+1$ is just $\beta_i - \beta_{i+1} =
\varphi_i - \delta$. 
\par
According to Eq.~(\ref{chain}), the dynamics of our supply network model can 
surprisingly be reduced to the dynamics of
a linear (sequential) supply chain, but with the new entities $i$ having the meaning of
``quasi-suppliers'' (analogously to ``quasi-species'' \cite{Eigen,Padgett}) defined by the linear combination
$\vec{\mu}(\tau)= \mathbf{T}^{-1} \vec{x}(\tau)$. 
This transformation makes it possible to define the bull-whip effect for arbitrary supply networks: It 
occurs if the amplitude $\mu_i^0$ is greater than the oscillation amplitude $\mu_{i+1}^0$ 
of the next downstream supplier $i+1$ and greater than the amplitude $h_i^0$ of the external forcing,
i.e. if $\mu_i^0/\max(\mu_{i+1}^0,h_i^0) > 1$. This {\em resonance effect} 
corresponds to the case of {\em convective instability}.
According to formula (\ref{help}), the bull-whip effect is particularly likely to appear, if the 
oscillation frequency $\alpha$ of the consumption rate is close to one of the resonance frequencies
$\omega_i$. 
\par\begin{figure}[htbp]
\begin{center}
\vspace*{-4mm}
\includegraphics[width=8.5cm]{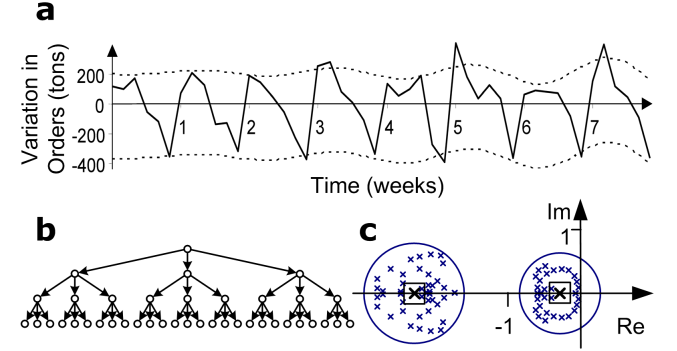} 
\vspace*{-4mm}
\end{center}
\caption[]{(a) Empirical example for a temporal 
amplification in the oscillation amplitude of the weekly order flow of a major company around
its average ($\gamma = 0.05$, $\omega = 2\pi/{\cal T}$ with ${\cal T} = 2$~weeks). (b) Example of a supply network 
(regular distribution network) with 
two degenerated real eigenvalues (see squares in the right subfigure). (c) The eigenvalues
$\lambda_{i,\pm}$ of the randomly perturbed supply network (crosses) are mostly complex
and  located within Ger\v{s}gorin's discs (large circles).}
\end{figure}
Depending on the respective network structure,
it can also happen that the oscillation amplitude of $\vec{\mu}(t)$ is amplified in the course of time
(see Fig.~1a). This case of {\em absolute instability} can occur 
if at least one of the eigenvalues $\lambda_{i,\pm}$ of the homogeneous equation
(\ref{chain}) resulting for $h_i = b_i = 0$ has a positive real part. The 
(up to) $2u$ eigenvalues 
\begin{eqnarray}
 \lambda_{i,\pm} 
= -\gamma_i \pm \sqrt{\gamma_i{}^2 - \omega_i{}^2} 
\label{omega2}
\end{eqnarray}
depend on the (quasi-)supplier $i$ and determine the temporal evolution of the amplitude of
deviations from the stationary solution. One can distinguish various interesting cases: (i) 
If the supply matrix $\mathbf{S}$ is symmetric, as for most mechanical or electrical oscillator networks,
all eigenvalues $J_{ii}$ are real. Consequently, if $\omega_i < \gamma_i$ (i.e. if $A$ is small enough),
the eigenvalues $\lambda_{i,\pm}$ of $\mathbf{M}$ are real and negative, corresponding to an
overdamped behavior. However, if $B$ is too small, the system behavior 
may be characterized by damped oscillations. (ii) Most natural and man-made supply networks have directed
links, and $\mathbf{S}$ is not symmetric. Therefore, some of the eigenvalues $J_{ii}$ will normally be complex,
and an overdamped behavior is untypical. The characteristic behavior is rather of oscillatory nature 
(although asymmetry does not always imply complex eigenvalues \cite{NATU}). 
For small values of $B$, it can even happen that the real part of an eigenvalue $\lambda_{i,\pm}$ 
becomes positive. This implies an amplification of the oscillations in time (until the oscillation amplitude is
limited by nonlinear terms). Surprisingly, this also applies to most upper triangular matrices, 
i.e. when no loops in the material flows exist. 
(iii) Another relevant case are linear (sequential) supply chains and regular supply networks (see, e.g., Fig.~1b). 
These are mostly characterized by degenerated zero eigenvalues $J_{ii}=0$ and Jordan
normal forms $\mathbf{J}$, i.e. non-vanishing upper-diagonal elements $J_{i,i+1}$. Sequential
supply chains and regular distribution systems belong to this case, which is characterized by the
two $u$-fold degenerate eigenvalues  $\lambda_{\pm} = -(1+B)/2 \pm \sqrt{(1+B)^2/4 - A}$
independently of the suppliers $i$. For small enough values $A<(1+B)^2/4$, these systems show overdamped
behavior, otherwise damped oscillations.
\par
Note that already very small perturbations in the network structure
can qualitatively change the dynamics of supply networks.
In case (iii), for example, they cause a cluster of complex eigenvalues around the two negative eigenvalues
$\lambda_{\pm}$ (if $A$ is small, see Fig.~1c). How can we explain this interesting observation? We may
use {\em Ger\v{s}gorin's theorem} on the location of eigenvalues \cite{Matrix}. 
Applying it to the perturbed matrix $\mathbf{B}_\eta = \mathbf{E} -\mathbf{S} + \eta \mathbf{P}$,
for small enough values of $\eta$, the corresponding eigenvalues should be
located within discs of radius $\eta \sum_{j} |P_{ij}^{(\eta)}|$ around the eigenvalues
$J_{ii}$ of the original matrix $\mathbf{E}-\mathbf{S}$. The
parameter $\eta$ with $0 < \eta \le 1$ allows to control the size of the perturbation.  
Moreover, $\mathbf{P}^{(\eta)} = \mathbf{R}_{\eta}^{-1} \mathbf{P} \mathbf{R}_{\eta}$, where
$\mathbf{R}_{\eta}$ is the matrix which transforms $\mathbf{B}_\eta$ to
a diagonal matrix $\mathbf{D}^{(\eta)}$, i.e. $\mathbf{R}_{\eta}^{-1} \mathbf{B}_\eta 
\mathbf{R}_{\eta} = \mathbf{D}^{(\eta)}$. (This assumes a perturbed matrix
$\mathbf{B}_\eta$ with no degenerate eigenvalues.) 
Similar discs as for the eigenvalues of $\mathbf{B}_\eta$
can be determined for the associated eigenvalues $\lambda_{i,\pm}^{(\eta)}$ 
of the perturbed $(2u\times 2u)$-matrix $\mathbf{M}_\eta$ belonging to 
the perturbed $u\times u$-matrix $\mathbf{B}_\eta$, see Eq.~(\ref{according}) and Fig.~1c.
\par
In summary, this contribution has shown how supply systems can be modeled by equations for 
material flows in networks. We could explain the empirically observed
bull-whip effect as a convective instability phenomenon 
based on a resonance effect and generalize it from linear (sequential) supply chains to arbitrary supply networks,
as these can be transformed on supply chains for ``quasi-suppliers''. However, the eigenvalues may become
complex. The dynamics of supply networks depends very sensitively on their topology and
two parameters $A$ and $B$ (which are related to derivatives of 
the ``control function'' of the delivery rates). In contrast to most systems of mechanical or electrical
oscillators, supply networks have directed links and are asymmetric.
Therefore, most supply networks are expected to show an damped or growing 
oscillatory behavior, even without loops. Although negative eigenvalues are found for regular distribution networks
and supply chains, already small perturbations of the network structure will cause a spectrum of
complex eigenvalues, i.e. a qualitative change in the dynamics. Our current
research focusses on the application to empirical input-output data and business cycles \cite{NATU},
on non-linear properties of supply networks, and applications to disaster management with a
lack of resources.
\par
We acknowledge partial support by SCA Packaging, 
DFG (research projects He 2789/5-1,6-1), ESF programme RDSES,
and KBN grant No. 5 P03A 025 20.

\end{document}